\documentclass[a4paper]{jpconf}
\usepackage{graphicx}

\usepackage{amsbsy,amssymb,latexsym,amsfonts,amsmath}

\allowdisplaybreaks

\newcommand{\al}{\alpha}
\newcommand{\be}{\beta}
\newcommand{\ga}{\gamma}
\newcommand{\de}{\delta}
\newcommand{\ep}{\epsilon}

\newcommand{\str}{{\rm STr}}
\newcommand{\alg}[1]{\mathfrak{#1}}
\newcommand{\el}{\nonumber}

\usepackage{xcolor}
\definecolor{light-gray}{gray}{0.9}

\begin{document}

\begin{flushright}
%\parbox{4cm}
ITP-UU-15/15 \\
KUNS-2579 
\end{flushright}

\title{Towards the gravity/CYBE correspondence beyond integrability 
-- {\it Yang-Baxter deformations of $T^{1,1}$} -- \footnote{This article is a brief review of the original paper \cite{CMY-T11} prepared for a proceeding of a talk given by T.M. at 
``The XXIIIth International Conference on Integrable Systems and 
Quantum symmetries (ISQS-23)"
held in Prague, Czech Republic, from June 23 till June 27, 2015.}}

\author{P. Marcos Crichigno}
\address{Institute for Theoretical Physics and Spinoza Institute, Utrecht University,\\
Leuvenlaan 4, 3854 CE Utrecht, The Netherlands.} 
\ead{p.m.crichigno@uu.nl}

\author{Takuya Matsumoto}
\address{Graduate School of Mathematics and Institute for Advanced Research,\\
Nagoya University, Nagoya 464-8602, Japan} 
\ead{takuya.matsumoto@math.nagoya-u.ac.jp}

\author{Kentaroh Yoshida}
\address{Department of Physics, Kyoto University, Kyoto 606-8502, Japan.} 
\ead{kyoshida@gauge.scphys.kyoto-u.ac.jp}

\begin{abstract}
Yang-Baxter sigma models, proposed by Klimcik and Delduc-Magro-Vicedo, 
have been recognized as a powerful framework for studying integrable deformations 
of two-dimensional non-linear sigma models. 
In this short article, as an important generalization, 
we review a {\it non-integrable} sigma model in the 
Yang-Baxter sigma model approach based on [arXiv:1406.2249]. 
%\href{http://arxiv.org/abs/1406.2249}{[arXiv:1406.2249]}. 
In particular, we discuss a
family of deformations of the 5D Sasaki-Einstein manifold $T^{1,1}$, instead of the standard
deformations of the $5$-sphere S$^5$\,. 
For this purpose, we first describe a novel construction of $T^{1,1}$ as a supercoset, and 
 provide a physical interpretation of this construction from viewpoint of the dual Klebanov-Witten field theory. Secondly, we consider a $3$-parameter deformation of $T^{1,1}$ by using classical
$r$-matrices satisfying the classical Yang--Baxter equation (CYBE).
The resulting metric and NS-NS two-form completely agree with the
ones previously obtained via TsT (T-dual -- shift -- T-dual)
transformations, and contain the Lunin-Maldacena background as a
special case. Our result indicates that what we refer to as the 
{\it gravity/CYBE(Classical Yang-Baxter Equation) correspondence} 
can be extended beyond integrable cosets.
\end{abstract}

%\tableofcontents

\section{Introduction and main results}

Ten-dimensional supergravity describes the low energy dynamics of string theory and has a rich mathematical structure. Finding explicit solutions to the supergravity field equations is an important task to better understand the low energy dynamics of string theory and to develop new examples of the AdS/CFT correspondence \cite{M,Witten:1998qj,Gubser:1998bc}, which is often limited to cases with a large amount of symmetry.
However, these equations are highly nonlinear and finding explicit solutions without a large amount of symmetry is not an easy task. 

One approach that has been quite successful in the past is to use known string theory duality transformations, such as T-duality \cite{Buscher}, as as solution-generating technique; starting from a known ``seed'' solution one applies this symmetry to generated a new solution.  A particular example are the so-called TsT transformations, which are a sequence of T-dualities (T) and coordinate shifts (s) leading to a continuous family of solutions \cite{LM,Frolov}. 

In a series of works in recent years \cite{MY-LM, MY-MR,MY-TsT,CMY-T11}, 
it has been observed that classical $r$-matrices, 
which are solutions to the classical Yang--Baxter equation ({\it CYBE} in short), 
characterize various deformations of type IIB supergravity backgrounds. One of the simplest examples is the correspondence between an Abelian $r$-matrix,
consisting of three Cartan generators of $\alg{su}(4)$\,, 
and the Lunin-Maldacena-Frolov backgrounds  \cite{LM,Frolov,MY-LM} 
obtained by TsT transformations. 
The gravity duals of the 4D non-commutative gauge theory \cite{HI,MR} 
can  also  be obtained by certain $r$-matrices \cite{MY-MR}. These observations have led us to propose a correspondence between classical $r$-matrices and deformed supergravity solutions, 
which we refer to as the {\it gravity/CYBE correspondence} \cite{MY-LM}. 

The reader may wonder how it is that $r$-matrices, 
which are purely algebraic objects,  can be used to study geometric objects such as gravity backgrounds. The notion which makes this possible is {\it Yang-Baxter sigma models}. These models, originally introduced by Klimcik \cite{Klimcik}, are integrable deformations of the principal 
chiral model based on an $r$-matrix satisfying the \textit{modified} classical Yang-Baxter equation ({\it mCYBE} in short).\footnote{
Quantum aspects of these models are discussed in 
\cite{Squellari,SST-lameta}.} 
These  were extended to sigma models defined on bosonic symmetric cosets \cite{DMV}, and applied to study  an integrable deformation of the AdS$_5\times$S$^5$ superstring \cite{MT} in \cite{DMV2,DMV3}. 
We note that the work \cite{DMV} is the generalization of 
{\it squashed sigma models} 
defined on the squashed S$^3$, and their integrable structures have been 
investigated in \cite{KYhybrid, KMY-QAA,KOY}.  
The deformed metric and the NS-NS two-form of the deformed 
AdS$_5\times$S$^5$ superstring were calculated in \cite{ABF}. 
However, the deformed background does not seem to be a supergravity solution for a generic value of the deformation parameter \cite{ABF-sugra}. 

Deformations of the AdS$_5\times$S$^5$ superstring by the CYBE (rather than by the mCYBE) were proposed in \cite{KMY-Jor} and the supergravity solution was constructed in \cite{KMY-SUGRA}. 
The generic formulation for  group manifolds and  bosonic cosets
was introduced in \cite{MY-PCM}. 
These works were inspired by a careful study of the three-dimensional 
case, namely {\it Schr\"odinger sigma models} 
\cite{KY-Sch,KY-exotic,KMY-Jor3d}. 
Although the CYBE is a particular case of the mCYBE, their solutions 
seem to be very different. 
One of the advantages of using the CYBE is the variety of classical $r$-matrices
and partial deformations of the target space that are possible. In the case of the mCYBE 
it seems difficult to find solutions, apart from the famous $r$-matrix
of the Drinfeld-Jimbo type \cite{Drinfeld1,Jimbo}. 
An important aspect of Yang-Baxter deformations is that they preserve the integrability of the system, $i.e.$, if the undeformed model is integrable, so is the deformed model. Thus, it is not surprising that the initial focus was on studying deformations of known integrable models such as AdS$_5\times$S$^5$ and four-dimensional Minkowski spacetime \cite{YB-min,YB-min-BKLSY}.\footnote{Integrable structures in AdS/CFT were discovered in \cite{MZ,BPR}. For a comprehensive review see \cite{review}.} 
The various deformed Lax pairs are summarized in \cite{KKSY-Lax}
and the expressions in terms of the local coordinates are presented. 
However, one may wonder whether integrability of the undeformed model is essential in the gravity/CYBE correspondence, or if this correspondence may hold beyond integrability.  If so, what are the resulting deformed backgrounds?

This question led us in \cite{CMY-T11} to consider deformations of the 
sigma model defined on the five-dimensional Sasaki-Einstein space known as 
$T^{1,1}$ \cite{CD} (see also \cite{Castellani:1983tb,R} for relevant work). 
As shown in \cite{BZ}, due to the appearance of chaotic behavior, this model is non-integrable. Thus, this is a good testing ground for possible generalizations of the gravity/CYBE correspondence in a non-integrable setting. Another important aspect of $T^{1,1}$ is that it admits a Sasaki-Einstein metric and, as a consequence, AdS$_5\times T^{1,1}$ is a good supergravity background. In fact, this was an early example \cite{KW}  of AdS/CFT with reduced supersymmetry. Thus,  the study of deformations of $T^{1,1}$ is also interesting from the holographic perspective. See \cite{Crichigno:2015pma} for a two-parameter deformation of metrics on $T^{1,1}$ in terms of gauged linear sigma models, and \cite{Benini:2015isa} for a one-parameter deformation and a study of supersymmetric localization of the two-dimensional sigma model on this space.

The first obstacle to apply the Yang-Baxter deformation to $T^{1,1}$ is how to describe the Sasaki-Einstein structure on the manifold by a coset.  Although it is well-known that $T^{1,1}$ \cite{CD,KW} is given by 
\begin{align}\label{usual coset}
\frac{SU(2)\times SU(2)}{U(1)_1}\,, 
\end{align}
the ``canonical'' coset metric on the space is not the Sasaki-Einstein metric we are interested in. Although this problem may be circumvented by a (rather artificial) rescaling of the vielbeins  (see App.\ A in \cite{CMY-T11} for  details and references), this is an unsatisfactory resolution from the point of view of the Yang-Baxter deformation; since the coset construction above does not make the full $SU(2)\times SU(2)\times U(1)$ symmetry of the space manifest, it is not clear how one may perform a  general (thee-parameter) Yang-Baxter deformation of the space.  As shown in \cite{CMY-T11} a novel construction which solves all these problems is given by the coset
\begin{align}
T^{1,1}=\frac{SU(2)\times SU(2)\times U(1)_R}{U(1)_1\times U(1)_2}\,.  
\label{T11-intro}
\end{align}
Although this coset is closely related to the one above, there is an important difference; despite involving only bosonic groups, it is in fact a {\it super-coset} in the sense that one must think of these bosonic groups as the bosonic part of a larger supergroup (this will be explained in detail below). Although this construction might seem at first unnatural, a clear advantage is that the full symmetry appears in the numerator of the coset and is therefore manifest. Furthermore, the fact that one should view this as a supercoset has a natural interpretation in terms of the ${\cal N}=1$ superconformal group $PSU(2,2|1)$ of the dual field theory.  

Having an appropriate description of the space as a coset,  the next step is to apply the Yang-Baxter deformation. As a simple example we consider deformations associated  with an Abelian classical $r$-matrix, given by
\begin{align}
r =\frac{1}{3} \hat\ga_1 L_3\wedge M+\frac{1}{3} \hat\ga_2  M\wedge K_3 
-\frac{1}{6} \hat\ga_3 K_3\wedge L_3\,, 
\end{align}
where $K_3$ and $L_3$ are the Cartan generators of two $SU(2)$'s respectively 
and $M$ is the $U(1)_R$ generator of the  
numerator in \eqref{T11-intro}\, and $\hat\gamma_1, \hat\gamma_2, \hat\gamma_3$ are three continuous parameters. Note that if the $U(1)_{R}$ was not present in the numerator, only a one-parameter deformation would be possible in this setting.  Although it is not guaranteed {\it a-priori} that such a deformation leads to a new supergravity background, it turns out that the deformed metric and the NS-NS two-form  agree exactly with the ones obtained by TsT-transformations \cite{CO} (in particular for
$\hat\gamma_1=\hat\gamma_2=0$\,, 
the deformed background reduces to the one given by Lunin-Maldacena \cite{LM}), and is therefore a good supergravity background. 
Thus, our result indicates that the Yang-Baxter sigma models based on the CYBE
can be applied beyond non-integrable cosets  
and the {\it gravity/CYBE correspondence} seems to extend to a wider class of  
supergravity backgrounds than one may have expected. 

This article is organized as follows. 
In Section \ref{sec:coset}, we discuss the new coset construction of $T^{1,1}$\,
 given in \eqref{T11-intro}. In Section \ref{sec:def}, we apply the Yang-Baxter deformation and show that the resulting backgrounds coincide with the TsT-transformed supergravity solutions, 
which corresponds to a $\beta$-deformation for the one-parameter case and to a 
$\ga_i$-deformation for the three-parameter case, respectively.  
Section \ref{sec:con} is devoted conclusions and discussions.

\section{Coset construction of $T^{1,1}$}
\label{sec:coset} 

\subsection*{$\S$ The metric of $T^{1,1}$}

Let us start by recalling some basic elements.  The five-dimensional space $T^{1,1}$  admits a Sasaki-Einstein metric, given by \cite{CD}
\begin{align}
ds^2_{T^{1,1}} &=\frac{1}{6}(d\theta_1^2+\sin^2\theta_1d\phi_1^2)
+ \frac{1}{6}(d\theta_2^2+\sin^2\theta_2d\phi_2^2) 
+\frac{1}{9}(d\psi+\cos\theta_1d\phi_1+\cos\theta_2d\phi_2)^2\,,    
\label{T11-metric}
\end{align}
where the angle variables $\theta_i\in [0,\pi), \phi_i\in [0,2\pi)$ with $i=1,2$
are the coordinates of the two two-spheres 
and the coordinate $\psi\in [0,4\pi)$ is along the $U(1)$-fiber.  The isometry of this manifold is $SU(2)\times SU(2)\times U(1)_R$\,, an important fact when studying general (Abelian) Yang-Baxter deformations below.

\subsection*{$\S$ Brief review of coset spaces} 

Before going into the details of the coset description of $T^{1,1}$, let us review 
some basics of general coset spaces and their metrics.  Let ${\cal M}$ be a homogeneous space, on which a Lie group $G$ is acting.   
For a fixed point of ${\cal M}$ we take a subgroup $H\subset G$ 
which preserves the reference point. 
Conceptually, $G$ is the global symmetry of the space ${\cal M}$
and $H$ is the local symmetry around the fixed point. 
Then, $\mathcal M$ corresponds to the coset space
\begin{align}
{\cal M}=\frac{G}{H}=\frac{\text{Global symmetry}}{\text{Local symmetry}}\,. 
\label{coset-ex}
\end{align}

A standard example of a coset space is the two-sphere S$^2$, whose global symmetry is 
$SO(3)\simeq SU(2)$\,. 
The residual symmetry around the north (or south) pole is $SO(2)\simeq U(1)$\,. 
This leads to the well-known coset description of the two-sphere; 
S$^2=SU(2)/U(1)$\,. 

To obtain the metric on ${\cal M}$ from the coset \eqref{coset-ex}, we need 
two ingredients. These are the Maurer-Cartan one-form
\begin{align}
A=g^{-1}dg\qquad \text{with}\qquad g\in G \,,
\end{align}
and the coset projector 
\begin{align}
P:\mathfrak{g} \to \mathfrak{g}/\mathfrak{h}\,, 
\end{align}
where $\alg{g}$ and $\alg{h}$ are the Lie algebras associated 
with the Lie group $G$ and $H$\,, respectively. 

Next, let us introduce the local coordinates $\{\theta^i\}$ 
through the parameterization of the group element $g\in G$\,,  
\begin{align}
g=\exp \left(\sum_{i=1}^{\dim \alg{g}}\theta^iT_i\right)
\qquad\text{with}\qquad 
T_i\in \alg{g}\,. 
\end{align}
Then, a metric $g_{ij}$ on $\mathcal M$ is obtained by 
\begin{align}
ds^2_{\cal M}={\rm Tr} [AP(A)] 
=\sum_{i,j=1}^{\dim {\cal M}}g_{ij}d\theta^i d\theta^j\,,  
\label{usual coset metric}
\end{align}
where the trace is taken over a particular representation of $\alg{g}$, 
which in our case will be the fundamental representation.

\subsection*{$\S$ Supercoset description of $T^{1,1}$}

Now we discuss the  (super)coset construction of  $T^{1,1}$ introduced in \cite{CMY-T11}. Unlike the usual  description (\ref{usual coset}), we take the numerator to be the \textit{full} isometry of the space and divide by an appropriate subgroup:
\begin{align}
T^{1,1}=\frac{SU(2)\times SU(2)\times U(1)_R}{U(1)_1\times U(1)_2}\,. 
\label{T11-coset}
\end{align}
We denote the generators of the two $SU(2)$'s by  $K_i, L_i$ ($i=1,2,3$), where $K_3$ and $L_3$ are the Cartan generators, and  we denote the generator of $U(1)_R$ by $M$.  The generators of $U(1)_i$ in the denominator are denoted by $T_i$ ($i=1,2$), which we choose to be embedded into the numerator by
\begin{align}
T_1=K_3+L_3\,,\qquad T_2=K_3-L_3+4M\,.
\label{embe}
\end{align}
We note that $T_{1}$ corresponds to the usual $U(1)_{1}$ in the standard coset description \cite{CD,KW}, while $T_{2}$  has been chosen appropriately in order to obtain the metric we are after.\footnote{
The manifolds $T^{p,q}$ with $p,q$ non-negative integers are  defined by the 
same coset $T^{p,q}\simeq [SU(2)\times SU(2)]/U(1)_1$, but with  $T_1=p K_3+q L_3$ (see \cite{CD} for details).  
This explains why the manifold is called $T^{1,1}$\,. }

%As mentioned above, the canonical coset metric does not lead to the correct (Sasaki-Einstein) metric \eqref{T11-metric}  without artificial rescaling of the vielbeins. In addition, the global symmetry of the space is is not manifest in this construction. MC: TOO REPETITIVE SO I COMMENTED THIS OUT 

The Maurer-Cartan one-form in our case is given by 
\begin{align}
A=g^{-1}dg \qquad \text{with}\qquad 
g\in SU(2)\times SU(2)\times U(1)_R\,, 
\end{align}
with the parameterization 
\begin{align}
g=\exp\bigl(\phi_1K_3+\phi_2L_3+2\psi M\bigr)
\exp\bigl(\theta_1K_2+(\theta_2+\pi)L_2 \bigr)\,. 
\end{align}
To define the coset projector explicitly, it is convenient to work with 
the fundamental representation. 
Since those of $\alg{su}(2)$ and $\alg{u}(1)$ are  $2\times 2$ 
and $1\times1$ matrices, respectively, 
the whole Lie algebra $\alg{su}(2)\oplus\alg{su}(2)\oplus\alg{u}(1)$
 is represented by a $5\times 5$ block matrix. 
We, however, adopt the following $(4|1)\times (4|1)$ {\it super}-matrix 
instead of the $5\times 5$ bosonic block matrix,  
\begin{align}
K_i=-\frac{i}{2}\left(
\begin{array}{cc|c} 
\sigma_i & 0 \,&\, 0 \\ 
0 & 0 \,&\, 0  \\ \hline  
0 & 0 \,&\, 0  
\end{array}\right),
\quad 
L_i=-\frac{i}{2}\left(
\begin{array}{cc|c} 
0 & 0 \,&\, 0 \\ 
0 & \sigma_i \,&\, 0  \\  \hline 
0 & 0 \,&\, 0  
\end{array}\right),
\quad 
M=-\frac{i}{2}
\left(
\begin{array}{cc|c} 
0 & 0 \,&\, 0 \\ 
0 & 0 \,&\, 0  \\ \hline  
0 & 0 \,&\, 1  
\end{array}\right), 
\label{sup-mat}
\end{align}
where $\sigma_i$ ($i=1,2,3$) are the Pauli matrices.\footnote{
The convention are given by ~~ 
$
%\begin{align}
\sigma_1=\begin{pmatrix}0 \,&\, 1 \\ 1 \,&\, 0\end{pmatrix}\,, \quad 
\sigma_2=\begin{pmatrix}0 & -i \\ i & 0\end{pmatrix}\,, \quad 
\sigma_3=\begin{pmatrix}1 & 0 \\ 0 & -1\end{pmatrix}\,. 
%\end{align}
$
}
As we shall  discuss in the next subsection, the appearance of the supermatrix is rather natural from the perspective of the full ten-dimensional AdS$_{5}\times T^{1,1}$ coset space. Since all off-diagonal blocks in \eqref{sup-mat} are zero, 
there are no fermionic elements. Thus, the essential difference in considering supermatrices, rather than bosonic ones, is that  in computing the metric one must use the {\it super}trace $\str$\,, rather than the $\Tr$, which is defined by 
\begin{align}
\str\left(\begin{array}{c|c} 
A & B  \\ \hline 
C & D  
\end{array}\right) \equiv \Tr(A)-\Tr(D)\,, 
\label{str}
\end{align}
where $A, D$ are the bosonic and $B, C$ are the fermionic block matrices. The representation given in \eqref{sup-mat} 
has the following normalization:   
\begin{align}
\str(K_aK_b)=\str(L_aL_b)=-\frac{1}{2} \de_{ab}\,, \qquad 
\str(MM)=\frac{1}{4}\,.  
\label{inpro}
\end{align}
Now, we define the coset projector as follows;
\begin{align}
P(x)=x-\frac{\str(T_1x)}{\str(T_1T_1)}T_1 - \frac{\str(T_2x)}{\str(T_2T_2)} T_2\,,    
\label{proj}
\end{align}
for $x\in \alg{su}(2)\oplus\alg{su}(2)\oplus\alg{u}(1)$\,. Explicit expressions for $T_1$ and $T_2$ follow  from 
\eqref{embe} and \eqref{sup-mat}. Finally, with this prescription, the metric (\ref{T11-metric}) is obtained by computing 
\cite{CMY-T11} 
\begin{align}
ds^2_{T^{1,1}}=-\frac{1}{3}\str[AP(A)]\,. 
\end{align}

\subsection*{$\S$ Why Supermatrices? }

One may wonder about the significance of the appearance the supertrace $\str$, rather than the usual trace. Although it would be interesting to address this question directly (and study possible generalizations to other spaces), here we offer an explanation of the case at hand based on AdS/CFT.  It is well known that backgrounds of the form AdS$_5\times X$, with $X$ a 5D Sasaki-Einstein manifold describe the gravity duals of certain ${\cal N}=1$ superconformal field theories.  The case $X=T^{1,1}$ was an early example proposed by Klebanov-Witten \cite{KW} 
(see \cite{Herzog:2002ih} for a review).  According to the AdS/CFT dictionary, the isometries of the internal manifold $X$ translate into global symmetries of the dual field theory. In the canonical case of $S^{5}$, the $SO(6)\sim SU(4)$ symmetry corresponds to the R-symmetry of the $\mathcal N=4$ dual field theory. In the case of $T^{1,1}$ the $U(1)_R$  part of the isometry corresponds to R-symmetry of the $\mathcal N=1$ dual field theory, while  the remaining $SU(2)\times SU(2)$ corresponds to the flavor symmetry of the field theory. Thus, these two symmetries should be physically distinguished. We recall that $U(1)_R$ is included in the bosonic subgroup of 
the ${\cal N}=1$ superconformal group $PSU(2,2|1)$ as follows, 
\begin{align}
SO(2,4)\times U(1)_R
=SU(2,2)\times U(1)_R
~\overset{\text{bosonic}}{\subset}~PSU(2,2|1)\,. 
\end{align}
Thus, from the perspective of the ten-dimensional space AdS$_5\times T^{1,1}$, the internal manifold $T^{1,1}$ in \eqref{T11-coset} appears naturally as the boconic part of the  supercoset\footnote{%%%%%%%%%%%%%%%%%%%%%%
Here it is worth recalling the more familiar case of AdS$_5\times$S$^5$\,, in which case the supercoset reads 
\begin{align}
\text{AdS}_5\times \colorbox{light-gray}{S$^5$} 
~ \simeq ~ \frac{SO(2,4)}{SO(1,4)}
\times 
\colorbox{light-gray}{$\displaystyle \frac{SU(4)_R}{SO(5)}$} 
~\overset{\text{bosonic}}{\subset}~ 
\frac{PSU(2,2\colorbox{light-gray}{$|4)$}}
{SO(1,4)\times \colorbox{light-gray}{$SO(5)$} }\,.
\end{align}
In this case, the R-symmetry is $SU(4)_R$ and is a part of the ${\cal N}=4$ superconformal algebra $PSU(2,2|4)$\,. 
}%%%%%%%%%%%%%%%%%%%%%%%
\begin{align}
\text{AdS}_5\times \colorbox{light-gray}{$T^{1,1}$} 
~~~ \simeq ~~~  &\frac{SO(2,4)}{SO(1,4)}
\times 
\colorbox{light-gray}{
$\displaystyle 
\frac{U(1)_R\times SU(2) \times SU(2)}
{U(1)_1\times U(1)_2}$} 
\nonumber 
\\
~\overset{\text{bosonic}}{\subset}~ 
&\frac{PSU(2,2\colorbox{light-gray}{
$|1)\times SU(2) \times SU(2)$}}
{SO(1,4)\times \colorbox{light-gray}{$U(1)_1\times U(1)_2$} }\,.
\end{align}
Here we have emphasized the $T^{1,1}$ part by the gray background.  Keeping in mind the above supercoset, the $(4|1)\times (4|1)$ supermatrix
in \eqref{sup-mat} 
may be justified by interpreting the matrix as a part of the bigger 
$(8|1)\times (8|1)$ supermatrix including the 
${\cal N}=1$ superconformal algebra as follows:
\begin{align}
\left(
\begin{array}{cc|c} 
\colorbox{light-gray}{$SU(2)$} & 0 & 0 \\ 
 0 & \colorbox{light-gray}{$SU(2)$} & 0  \\ \hline  
 0 & 0 & \colorbox{light-gray}{$U(1)_R$}  
\end{array}
\right)
\quad \hookrightarrow  \quad 
\left(
\begin{array}{ccc|c} 
SU(2,2) & 0 & 0 & \overline{F}{}^A \\ 
0 & \colorbox{light-gray}{$SU(2)$} & 0 & 0 \\ 
0 & 0 & \colorbox{light-gray}{$SU(2)$} & 0  \\ \hline  
F_A & 0 & 0 & \colorbox{light-gray}{$U(1)_R$}  
\end{array}
\right)\,.    
\label{big supermatrix}
\end{align}
In the larger supermatrix on the right of (\ref{big supermatrix}) the ${\cal N}=1$ superconformal algebra
$PSU(2,2|1)$ is sitting at the four corners, together with the eight supercharges 
$F_A, \overline{F}{}^A$ ($A=1,\cdots,4$).  
From this supermatrix one easily recognizes that the lines separating the $\mathbb{Z}_2$-grading
of the $(4|1)\times (4|1)$ supermatrix on the left arises  from the structure of the  ${\cal N}=1$ superconformal algebra $PSU(2,2|1)$ on the right. Thus, the appearance of supermatrices (and with this particular grading) is illuminated by AdS/CFT.

\section{Deformations of $T^{1,1}$ as Yang-Baxter sigma models}
\label{sec:def} 

\subsection*{$\S$ Yang-Baxter sigma models}

Now that we have discussed the coset  \eqref{T11-coset} in detail,
we apply the Yang-Baxter sigma model approach to study deformations of this space. We will use a classical $r$-matrix \cite{MY-PCM,KMY-Jor}. 

The Lagrangian of the Yang-Baxter sigma model on $T^{1,1}$ is defined by 
\begin{align}
L=\frac{1}{3}(\ga^{\al\be}-\ep^{\al\be})\str 
\left[A_\al P\frac{1}{1-2R_g\circ P} A_\be\right] \,, 
\label{Lag}
\end{align}
where the coset projector $P$ given in \eqref{proj}  
and the flat metric on the world-sheet $\ga^{\al\be}$ and 
the anti-symmetric tensor $\ep^{\al\be}$ are 
normalized as $\ga^{\al\be}={\rm diag}(-1,+1)$ and $\ep^{\tau\sigma}=1$\,.   
The {\it dressed} operator $R_g$ with 
$g\in SU(2)\times SU(2)\times U(1)_R$ is defined by 
\begin{align}
R_g(x)\equiv g^{-1}R(gxg^{-1})g\,,  
\end{align} 
where the linear $R$-operator satisfies the CYBE (classical Yang-Baxter equation) 
\begin{align} 
[R(x),R(y)]-R([R(x),y]+[x,R(y)])=0 
\end{align}
for $x,y\in \alg{su}(2)\oplus \alg{su}(2)\oplus\alg{u}(1)_R$\,, 
rather than the mCYBE (modified classical Yang-Baxter equation) \cite{DMV2}\,. 

The deformed projected current in the Lagrangian \eqref{Lag}
\begin{align}
P(J_\be):=P\frac{1}{1-2R_g\circ P} A_\be
\end{align}
can be obtained by solving the equation
\begin{align}
(1-2P\circ R_g )P(J_\be) = P(A_\be) \,. 
\end{align}

\subsection*{$\S$ Classical $r$-matrix and the linear $R$-operator}

Deformations of Yang-Baxter sigma models \eqref{Lag} are 
characterized by the linear $R$-operator, and reduces to the undeformed model for $R=0$\,. 
In the case of AdS$_5\times$S$^5$\,, it was revealed in \cite{MY-LM}
that the $R$-operators which  
non-trivially act on the Cartan generators of $\alg{su}(4)$ yield the $\beta$- or $\hat{\ga}_i$-deformed backgrounds \cite{LM,Frolov}. As a simple generalization of the case S$^5$ case, 
here we also consider the Abelian $r$-matrix consisting of 
the three Cartan generators,\footnote{
In principle, the $R$-operator does not have to satisfy the (m)CYBE
since the undeformed model is no longer integrable. 
}
\begin{align}
r =\frac{1}{3} \hat\ga_1 L_3\wedge M+\frac{1}{3} \hat\ga_2  M\wedge K_3 
-\frac{1}{6} \hat\ga_3 K_3\wedge L_3\,, 
\label{abe-3para}
\end{align}
with three deformation parameters $\hat\ga_i$ ($i=1,2,3$). 
The wedge stands for the skew-symmetric tensor product 
$x\wedge y\equiv x\otimes y-y\otimes x$
and $K_3$ and $L_3$ are the Cartan generators of two $\alg{su}(2)$'s
and $M$ is the $\alg{u}(1)_R$ generator \eqref{sup-mat}.  

In general, from a tensorial classical $r$-matrix such as in \eqref{abe-3para}, 
the linear $R$-operator is obtained by tracing out the second entries;
\begin{align}
&R(x)=\str_2[r(1\otimes x)]\equiv \sum_i \bigl(a_i\str(b_i x)-b_i\str(a_i x)\bigr)\,,  
\label{linearR} 
\\
\text{with}\quad &r=\sum_i a_i\wedge b_i
=\sum_i (a_i\otimes b_i-b_i\otimes a_i)\,. \el 
\end{align}
Here it is noted that, as a special feature of $T^{1,1}$\,,  
we shall use the supertrace $\str$ in \eqref{str} instead of the usual one.

Using the inner products in \eqref{inpro}, the explicit action of 
the linear $R$-operator associated with the 
classical $r$-matrix \eqref{abe-3para}\, reads  
\begin{align}
R \begin{pmatrix} K_3 \\ L_3 \\ M \end{pmatrix}
=-\frac{1}{12}\begin{pmatrix} 
0& \ga_3 & 2\ga_2 \\
-\ga_3& 0 & -2\ga_1 \\
\ga_2 & -\ga_1 & 0 
\end{pmatrix} \begin{pmatrix} K_3 \\ L_3 \\ M\end{pmatrix}\,, 
\quad 
R(\text{others})=0\,.  
\label{linear-R}
\end{align}

\subsection*{$\S$ Deformed backgrounds}

Plugging the linear $R$-operator \eqref{linear-R} into the Yang-Baxter sigma model
\eqref{Lag}, one can read off 
the deformed metric and the NS-NS two-from \cite{CMY-T11}\,;  
\begin{align}
ds^2 &= G(\hat\ga_1,\hat\ga_2,\hat\ga_3)\Bigl[
\frac{1}{6}\sum_{i=1,2}(G(\hat\ga_1,\hat\ga_2,\hat\ga_3)^{-1}d\theta_i^2+\sin^2\theta_i d\phi_i^2) \el \\
&\quad 
+\frac{1}{9}(d\psi+\cos\theta_1d\phi_1+\cos\theta_2d\phi_2)^2 
+\frac{\sin^2\theta_1\sin^2\theta_2}{324}(\hat\ga_3 d\psi+\hat\ga_1 d\phi_1+\hat\ga_2 d\phi_2)^2 \Bigr]\,, 
\\
B_2 &=G(\hat\ga_1,\hat\ga_2,\hat\ga_3)\Bigl[
\Bigl\{\hat\ga_3 \Bigl(\frac{\sin^2\theta_1\sin^2\theta_2}{36}
+\frac{\cos^2\theta_1\sin^2\theta_2+\cos^2\theta_2\sin^2\theta_1}{54}\Bigr)\el \\
&\hspace{33mm}
-\hat\ga_2 \frac{\cos\theta_2\sin^2\theta_1}{54}
-\hat\ga_1 \frac{\cos\theta_1\sin^2\theta_2}{54} 
\Bigr\}d\phi_1\wedge d\phi_2 \el\\
&\quad 
+\frac{(\hat\ga_3\cos\theta_2-\hat\ga_2)\sin^2\theta_1}{54}d\phi_1\wedge d\psi 
-\frac{(\hat\ga_3\cos\theta_1-\hat\ga_1)\sin^2\theta_2}{54}d\phi_2\wedge d\psi 
\Bigr]\,, 
\end{align}
where the scalar function $G$ is defined by 
\begin{align}
G(\hat\ga_1,\hat\ga_2,\hat\ga_3)^{-1}&\equiv1
+\hat\ga_3^2 \Bigl(\frac{\sin^2\theta_1\sin^2\theta_2}{36}
+\frac{\cos^2\theta_1\sin^2\theta_2+\cos^2\theta_2\sin^2\theta_1}{54}\Bigr)
+\hat\ga_2^2 \frac{\sin^2\theta_1}{54} \el \\
&\qquad 
+\hat\ga_1^2 \frac{\sin^2\theta_2}{54} 
-\hat\ga_2\hat\ga_3 \frac{\sin^2\theta_1\cos\theta_2}{27} 
-\hat\ga_3\hat\ga_1 \frac{\sin^2\theta_2\cos\theta_1}{27} \,. 
\end{align}
These  expressions for the metric and NS-NS two-form coincide exactly with the ones obtained in \cite{CO}
via TsT-transformations.\footnote{In the special case $\hat\ga_1=\hat\ga_2=0$ and $\hat\ga_3\neq 0$ one obtains the Lunin-Maldacena background \cite{LM}.} Thus, we have shown by means of an explicit example that the {\it gravity/CYBE correspondence} holds beyond the strict case of integrable backgrounds.

\section{Conclusions and discussions}
\label{sec:con} 

The main question addressed in this short article is whether the 
{\it gravity/CYBE correspondence} \cite{MY-LM,MY-MR,MY-TsT} 
can be extended beyond the integrable case. The answer seems to be yes, as first realized in \cite{CMY-T11} and reviewed here. 

To recapitulate, we have shown this to be the case by means of  an explicit example, namely the  5D Sasaki-Einstein space  $T^{1,1}$, which was shown to be non-integrable in \cite{BZ,AKKY-chaos}. Along the way to showing this we had to develop a novel (super)coset construction for this space, namely \eqref{T11-coset}. As discussed, unlike the standard coset construction, this prescription leads directly to the Sasaki-Einstein metric on the space and is specially suited to applying the  Yang-Baxter deformation. Although the origin of the supermatrix construction is not entirely clear from  a purely mathematical standpoint, it is a rather natural construction from the viewpoint of the ${\cal N}=1$ superconformal symmetry of the dual field theory.  It would be interesting to study the mathematical origin of this construction, as well as possible generalizations of it to other coset spaces. 

Next, we have applied the Yang-Baxter sigma model deformation with a general Abelian $r$-matrix \eqref{abe-3para}, showing that the resulting deformed metric and the NS-NS two-form are in complete agreement with the deformed supergravity backgrounds
obtained by TsT-transformations, corresponding to  $\beta$- or $\hat{\ga}_i$-deformations \cite{LM,CO}. This establishes that integrability does not seem to be an essential feature in the gravity/CYBE correspondence. 
At the moment, it is not clear to the authors if this relation is specific to $T^{1,1}$ or if it may extend to general non-integrable backgrounds.
Thus, it would certainly be interesting to explore further examples.  One may consider for example deformations of $T^{1,1}$ by non-Abelian $r$-matrices, or possible extensions to the general family of Sasaki-Einstein manifolds such as $Y^{p,q}$\,, which are referred as to {\it cohomogeneity $1$}.\footnote{Although this may seem possible conceptually, the general case is technically much more difficult.} At a more fundamental level, it would be very interesting to have a better understanding in general of the role of the CYBE providing deformations of non-integrable backgrounds. 

It would also be interesting to study whether the supercoset construction described here, and its Yang-Baxter deformations, can be realized in terms of generalized $\mathcal N=(2,2)$ gauged linear sigma models, along the lines of \cite{Benini:2015isa,Crichigno:2015pma}. 

Other deformations of sigma models which have been proposed and 
are actively studied are \cite{HMS1, lambda}.\footnote{
For the further references of the Yang-Baxter sigma models, 
for instance, see \cite{HRT,AvT,AdLvT,Tongeren}.} 
One of them is referred as to {\it $\lambda$-deformations} \cite{lambda}. 
The relation between $\eta$-deformations, which are based on 
 classical $r$-matrices of the Drinfeld-Jimbo type,  
and  $\lambda$-deformations was considered by 
\cite{Vicedo, HT-lameta,SST-lameta}, where it was explained 
that these are {\it Poisson-Lie T-dual}. 
The integrated picture is also captured from the perspective of 
{\it the ${\cal E}$-models} by \cite{Klimcik-E}.

\section*{Acknowledgments}

We are grateful to Gleb Arutyunov, Riccardo Borsato, Martin Ro\v cek 
and Stefan Vandoren for valuable comments and discussions 
for the work \cite{CMY-T11}. 
P.M.C. is supported by the Netherlands Organization for Scientific Research (NWO) under the VICI Grant 680-47-603. This work is part of the D-ITP consortium, a program of the NWO that is funded by the Dutch Ministry of Education, Culture and Science (OCW).
The work of K.Y. is supported by Supporting Program for 
Interaction-based Initiative Team Studies 
(SPIRITS) from Kyoto University and 
by the JSPS Grant-in-Aid for Scientific Research (C) No.15K05051.
This work is also supported in part by the JSPS Japan-Russia Research 
Cooperative Program 
and the JSPS Japan-Hungary Research Cooperative Program.

\end{document}